# The Influence and Relationship between Computational Thinking, Learning Motivation, Attitude, and Achievement of Code.org in K-12 Programming Education

WAN CHONG CHOI[a]* & IEK CHONG CHOI[b]

[a] *Department of Computer Science, Illinois Institute of Technology, USA;*
[b] *School of Education, City University of Macau, Macao SAR, China*

Corresponding author: Wan Chong Choi (e-mail: wchoi8@hawk.iit.edu)

10 West 31st Street, Stuart Building, Chicago, IL 60616, USA

**ORCID**

WAN CHONG CHOI https://orcid.org/0000-0002-8415-6998

IEK CHONG CHOI https://orcid.org/0009-0000-0241-7166

# The Influence and Relationship between Computational Thinking, Learning Motivation, Attitude, and Achievement of Code.org in K-12 Programming Education

ABSTRACT

This study investigated the impact of Code.org's block-based coding curriculum on primary school students' computational thinking, motivation, attitudes, and academic performance. Given their importance in today's digital world, the study involved 20 students and adopted a design to address the challenges in nurturing programming skills in young learners.

The Programming Computational Thinking Scale (PCTS) was utilized to evaluate computational thinking. Student motivation was measured using the Instructional Materials Motivation Survey (IMMS), while the Attitude Scale of Computer Programming Learning (ASCOPL) assessed students' attitudes toward programming. The Programming Achievement Test (PAT) was employed to gauge students' performance in programming tasks.

The results showed significant enhancements in computational thinking, increased motivation for learning, positive attitudes toward programming, and remarkable academic achievements in programming tasks. Moreover, strong positive correlations were identified among computational thinking skills, learning motivation, attitudes, and student achievement on Code.org. These findings underscored the interrelatedness of these factors in influencing programming learning outcomes.

Further analysis through ANOVA revealed significant differences in achievement levels in computational thinking (specifically Computational Concepts and Perspectives) and learning attitudes (Willingness, Negativity, Necessity), along with motivational factors (Attention, Relevance, Confidence). This underscored the complex relationship between these factors in contributing to programming success.

Our study contributed by thoroughly exploring the interconnectedness of computational thinking, learning motivation, attitudes, and their collective impact on programming achievement within block-based programming in primary school. This work illuminates the importance of these factors in fostering a conducive learning environment for programming, addressing a notable gap in existing literature.

**Introduction**

In the digital era, the development of programming skills is increasingly recognized as a crucial component of education, particularly for younger students. In Macao, integrating programming curricula in primary schools is a testament to this trend, highlighting the importance of cultivating skills such as computational thinking from an early age (Wing, 2006). Computational thinking, the ability to analyze and redesign systems to solve problems using computer-based solutions, is at the core of programming education. However, fostering these skills in students is not without its challenges. Research has shown that while students can replicate programming tasks, their understanding of the conceptual underpinnings often remains limited, resulting in difficulties when they attempt to program independently (Tsou, 2014).

The learners' attitude and motivation are pivotal factors influencing the effectiveness of programming education (Erodogan et al., 2008). Positive learning attitudes substantially impact educational achievement (Hwang et al., 2012) (Lai et al., 2012). Traditional programming curricula, often based on text-based languages, present substantial barriers to beginners due to their complex syntax and logical structures, reducing interest and lower learning attitudes (Felleisen et al., 2004). This decline in interest can also adversely affect students' learning attitude and motivation, a key driver in sustaining engagement and perseverance in challenging subjects like programming (Agapito et al., 2017).

Addressing these issues, the emergence of block-based programming languages, exemplified by platforms like Code.org, offers a more accessible and engaging approach to learning programming. These platforms simplify the programming process through visual elements, eliminating the need to understand intricate syntax, thus making it easier for a broader range of students to learn coding logic (B. A. Myers, 1986) (Weintrop, 2019). Incorporating gamification elements in Code.org has shown the potential to enhance learning motivation. By engaging students in game-based challenges, these platforms can significantly improve both problem-solving skills and computational thinking (Agapito et al., 2017) (Kim & Kim, 2017).

Despite the advancements in block-based programming and its potential to positively impact learning motivation and attitude, there is a noticeable gap in research, especially concerning its influence on computational thinking, learning motivation, and learning attitude among primary school students. Previous studies have predominantly focused on text-based programming, with a limited exploration of the multifaceted impact of block-based programming in primary education (Maryono et al., 2022). The research aimed to bridge the existing gap by exploring how Code.org impacts learning motivation, attitudes, computational thinking, and academic achievement within the scope of programming instruction. The subsequent questions were considered in this study:

1) How does learning through Code.org impact students' computational thinking skills?
2) What effect does Code.org have on students' learning motivation?
3) How do their experiences with Code.org influence students' learning attitudes?
4) What are the academic achievements of students who study via Code.org?
5) What kind of relationship between computational thinking, learning motivation, learning attitude, and academic achievement in Code.org?

**Literature Review**

*Computational Thinking*

Computational thinking involves using specific cognitive processes to address complex problems. It is about breaking these problems into smaller, more manageable parts, which is crucial for effective problem-solving. This approach includes formulating detailed algorithms and step-by-step procedures, thoroughly analyzing data, designing and implementing solutions, and applying logical and computational concepts to various situations. It is a systematic way of thinking that helps understand and solve complex issues using fundamental computing principles (Selby & Woollard, 2013). Computational thinking emphasizes the ability to think algorithmically and systematically, enabling individuals to approach problems in a structured and efficient manner, whether with or without the assistance of computers (Wing, 2008).

Computational thinking and programming education are closely related, as programming provides a practical application of computational thinking concepts. Programming involves implementing algorithms and instructions using programming languages and translating computational thinking into executable code (Bers et al., 2014). Integrating computational thinking into programming education aims to develop students' computational skills, such as algorithmic thinking, logical reasoning, abstraction, problem decomposition, and debugging strategies. Programming education allows students to apply computational thinking concepts, fostering their understanding of computational processes and enhancing their problem-solving abilities (Werner et al., 2012).

In K-12 education, computational thinking has become a vital competency for students to navigate an increasingly digital world. Efforts have been made to introduce computational thinking into the curriculum, requiring cooperation with the computer science education community. However, teaching computational thinking in K-12 education demands systemic alterations, active teacher involvement, and the creation of substantial resources (Barr & Stephenson, 2011). Research in this area has focused on investigating the development of students' computational thinking skills, exploring appropriate computational thinking models, assessing computational thinking performance, and designing instructional materials that integrate computational thinking into various subjects and grade levels (Weintrop et al., 2016) (Chen et al., 2017) (Tang et al., 2020).

A computational thinking model has been proposed by the Massachusetts Institute of Technology's (MIT) Lifelong Kindergarten Group. (Brennan & Resnick, 2012) to provide a comprehensive understanding of computational thinking and include the following elements:

Computational Concepts: In programming education, students are required to understand the seven foundational programming concepts, encompassing sequences, parallelism, conditions, events, data, loops, and operators.

Computational Practices: To solve problems through programming effectively, applying and practicing the computational concepts learned is essential for learners. Debugging and problem-solving help them get a better understanding of computational thinking concepts.

Computational Perspectives: Fundamentally, connection, expression, and questioning constitute the three essential dimensions. They are associated with learners' creativity, representational skills, communication, collaboration, problem-understanding, and inquiry capabilities.

Moreover, numerous scholars have identified a significant relationship between computational thinking and programming or computing concepts, leading to a focus on interventions centered around programming activities. They measure students' programming skills as an indicator of their computational thinking proficiency. This relation is well-demonstrated in platforms like Scratch, a block-based programming environment developed at MIT for creating interactive media. Scratch has facilitated a series of studies defining computational concepts in alignment with programming concepts (Grover et al., 2015) (Román-González et al., 2017) (Pugnali et al., 2017).

*Learning Motivation*

Motivation is a critical theoretical construct that illuminates human behavior by stimulating individuals to respond and fulfill their needs (Aldrich, 2009) (Bayliss & Strout, 2006). Specifically, motivation encompasses initiating, directing, and sustaining goal-directed behavior, directing individuals to take action toward achieving a goal or fulfilling a need or expectation. Scholars have identified motivation as a critical factor influencing students' learning outcomes (Huang & Hew, 2016) (Law et al., 2019). However, the high failure and dropout rates in introductory computer programming courses can be attributed to students' lack of motivation to engage (Kazimoglu, 2020). In dealing with this concern, gamification has been proposed as a potential motivational strategy in programming teaching, as it offers interactive, entertaining, and creative experiences (Aldrich, 2009). Gamification aims to motivate students to engage in learning activities, reduce the gap between theory and practice, and bridge the divide between abstract concepts and concrete activities.

The ARCS model, which comprises four dimensions: Attention, Confidence, Relevance, and Satisfaction, was used to evaluate how programming with Code.org affects students' learning motivation. Keller proposed the model of ARCS (Keller, 1979) according to the framework of expectancy-value theory. Each of the four dimensions within the ARCS model is crucial in enhancing learning motivation as follows.

Attention: This dimension focuses on capturing students' interest and curiosity. It involves strategies that make the learning experience engaging and attention-grabbing (Plass et al., 2015). In the context of programming with Code.org, this can include integrating interactive elements, attractive graphics, or engaging challenges to spark students' curiosity and keep them entertained during the learning process.

Relevance: Relevance in education, mainly when teaching programming through platforms like Code.org, focuses on making the content meaningful and applicable to students' lives. It involves highlighting the importance of the learning material by emphasizing its real-world applications. Educators can significantly enhance students' motivation by demonstrating how coding skills are relevant in various contexts (Gee, 2003). Students are more likely to engage when they understand the practical significance of their learning.

Confidence: The Confidence dimension focuses on building students' self-assurance and belief in their abilities to master the material (Bandura et al., 1999). When using Code.org for programming education, providing a supportive learning environment, offering gradual challenges with opportunities for success, and giving constructive feedback can boost students' Confidence in their programming skills. Confidence is essential for sustained motivation and a sense of accomplishment.

Satisfaction: It refers to the overall enjoyment and positive learning experience (Mirvis, 1991). It encompasses the fulfillment that comes from completing tasks or achieving goals. In the context of programming with Code.org, designing rewarding learning experiences, celebrating students' achievements, and creating a sense of accomplishment can enhance satisfaction. When students find satisfaction in their learning journey, they are more likely to remain motivated and eager to continue their programming education.

In this paper, we facilitated students' learning of programming knowledge by having them participate in level-based gaming challenges with Code.org to enhance their performance effectively across the various dimensions of the ARCS model.

*Learning Attitude*

When examined from a psychological standpoint, attitudes encompass three dimensions: Affective, Behavioral Intention, and Cognition. The Attitude ABCs is the collective term (D. G. Myers, 1993). Wang et al. (W. K. Wang et al., 1995) assert that attitudes stemming from learning significantly influence behavioral reactions to particular elements and represent the structured internal psychological states of preferences and aversions. Furthermore, described as the pathway, learning supports the advancement of cognition, behavior, and emotion (H. W. Wu, 2013).

Experts offered various interpretations of learning attitudes. Nga (Ngan, 1980) viewed it as an environment influenced by an individual's psychological tendencies, including curriculum, processes, and methods. Zhu (Zhu, 1986) identified six essential traits of a positive learning attitude: defined goals, keen interest, innovative approaches, comprehension of materials, ongoing engagement, and practical resource use. Wang (K. X. Wang, 1996) saw these attitudes as enduring psychological responses involving cognitive, emotional, and behavioral aspects. Chang (Chang, 1996) defined learning as a behavior-modifying process shaped by practice and attitude. Finally, Lee (Li, 2014) expanded this view to a holistic approach, integrating cognitive, behavioral, and emotional elements essential for knowledge acquisition.

The significance of students' attitudes, particularly in programming, could not be overstated, as these attitudes profoundly impacted their learning effectiveness (Liaw, 2008). Negative attitudes towards learning impeded Confidence and new information assimilation, adversely affecting performance. Conversely, a positive outlook encouraged receptiveness to novel programming concepts (Yang et al., 2018). Tan (Tan et al., 2009) highlighted that students' perceived programming as challenging diminished their interest, further influencing their learning outcomes.

*Block-Based Programming Language*

To make programming more accessible, dragging and dropping take precedence over text input in block-based programming languages. Additionally, it embraced interaction and offered visual support to facilitate students in learning programming (Malan & Leitner,

2007). Code.org was the platform employed as the block-based programming language in this study.

Code.org serves as a platform engaged in block-based programming and is known for its commitment to simplifying the process of learning programming, particularly for novice learners, such as students in K-12 education. The platform's drag-and-drop interface replaces traditional text-based coding, reducing the cognitive load associated with coding syntax and making programming more accessible and less intimidating (Noone & Mooney, 2018). Visual representation is a core element, with a library of colorful and intuitive code blocks representing various programming constructs, such as loops, conditionals, and variables. This visual approach may help students comprehend programming concepts more quickly.

Code.org emphasizes interactivity, allowing students to actively engage with code by running their programs in real-time, as shown in Figure 1. This hands-on experience lets learners see the immediate results of their code, motivating them to continue exploring and learning. Furthermore, they can experiment more quickly when they see the outcomes of what they created (Noone & Mooney, 2018). This platform provides interactive coding challenges and step-by-step tutorials, guiding learners through a gradual progression of more complex coding concepts (X.-M. Wang et al., 2017). The curriculum enhances problem-solving and computational thinking skills, reducing the need for intense memorization (Kalelioğlu et al., 2014). This approach ensures a smooth learning curve and enables students to utilize their programming knowledge and skills better.

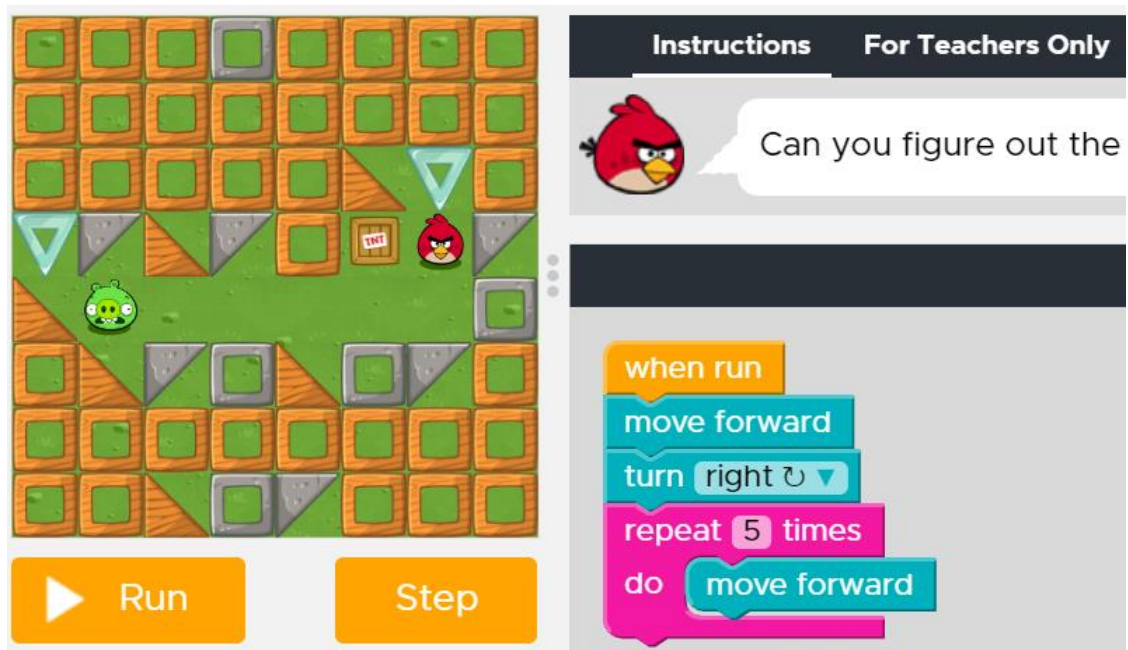

Figure 1. Interface of Code.org

Game-based learning is another highlight of Code.org, incorporating game elements and interactive puzzles into its curriculum. This gamified approach makes learning programming enjoyable and encourages students to complete challenges, earn rewards, and advance through lessons. The platform offers diverse coding courses for students of various ages and skill levels. From introductory courses like "Computer Science Fundamentals" for elementary students to more advanced courses like "AP Computer Science Principles", learners can choose the path that aligns with their capabilities and interests.

Code.org also places a strong emphasis on accessibility and inclusivity. It is designed to be accessible to many learners, including those with disabilities. The platform provides various resources and translations to make coding education available to a global audience. For educators, Code.org offers help and support to facilitate teaching programming in the classroom. This includes lesson plans, teacher guides, and professional development opportunities, ensuring teachers can effectively implement coding education in their courses.

The platform aligns its curriculum with national and international educational standards, making Code.org courses relevant and complementary to traditional K-12 education. It fosters community and collaboration among learners, allowing students to share their projects, learn from peers, and collaborate on coding challenges. Code.org's commitment to simplifying programming and making it accessible to learners of all backgrounds has made it a valuable resource for K-12 education and beyond, enhancing the educational experience and preparing students for the digital age.

**Methodology**

*Research Design*

The investigation adopted a quasi-experimental approach within a single-group framework. Utilizing a one-group pretest-posttest design (Büyüköztürk et al., 2008), the research assessed the impact of Code.org on learning motivation, attitudes, achievement, and computational thinking. This design was chosen to provide insights into cause-and-effect relationships and trace the evolution of the experiment's impact. Such a methodology effectively reveals the study's progress and outcomes, showcasing the learning and cognitive development changes throughout the teaching experiment.

*Participants*

This investigation was carried out at a Macao primary school where the first author is employed. It involved a participant group of 20 students in the third grade at a primary school aged 9 to 10 years old.

*Teaching Materials*

The teaching schedule spanned three weeks, encompassing a total of 18 lessons. During this period, there were 6 lessons each week, with each lesson lasting 40 minutes.

Instruction in block-based programming took place using the Code.org website. The instructional material consists of Course B from the Code.org Computer Science Fundamentals curriculum. The Code.org activities concentrated on instilling the basics of algorithms, highlighting conditions, sequence (Figure 2), nested loops, and loops (Figure 3). Learners will engage in more intricate unplugged assignments, and the course will end with puzzles. Students will receive instruction in programming fundamentals, skills for teamwork, critical and investigative thinking abilities, and perseverance in the face of challenges.

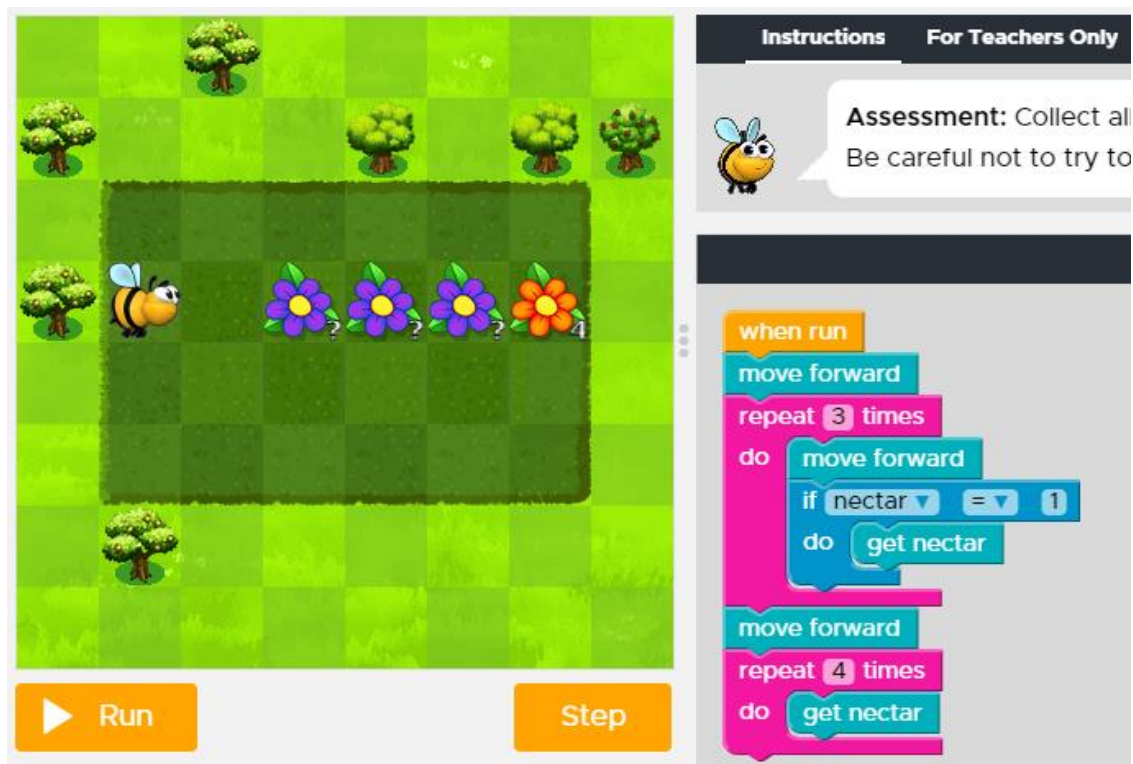

Figure 2. The conditions program of Code.org

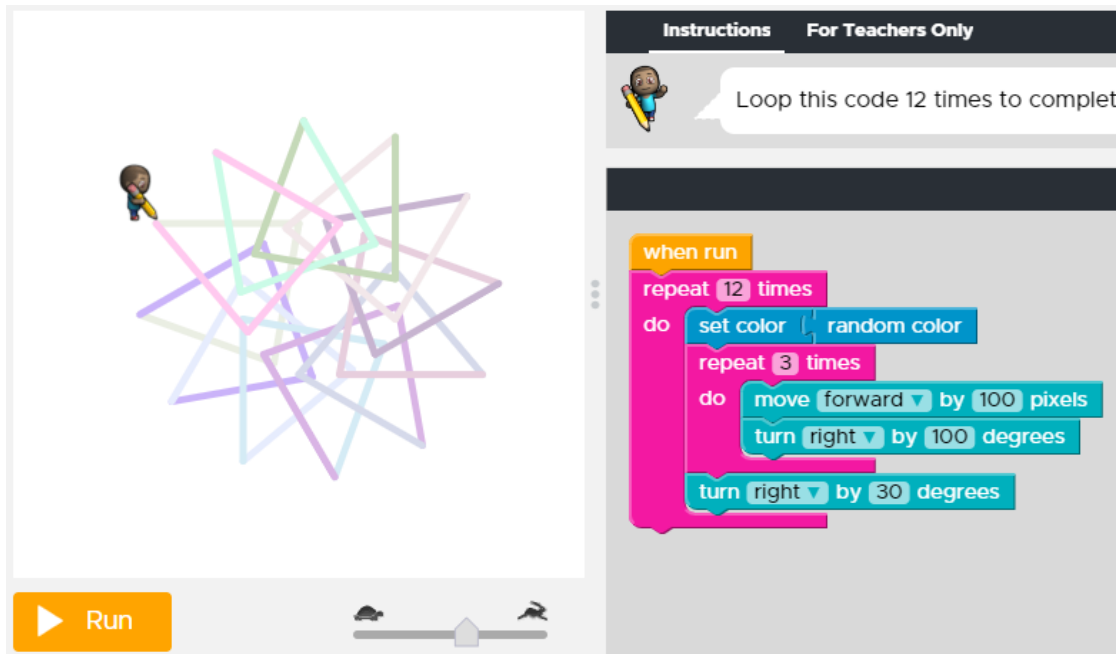

Figure 3. The nested loops program of Code.org

*Research Instrument*

This study used the different questionnaires as quantitative measures. The implementation's impact was assessed by analyzing the differences among the three questionnaires. The study's structure is outlined in Table 1.

Table 1. Design of pretest-posttest

| N | Pre-Test | Teaching Experiment | Post-Test |
|---|---|---|---|
| 20 | $O_1$ $O_2$ $O_3$ | X | $O_4$ $O_5$ $O_6$ |

X: The group experienced the experimental treatment with the aid of Code.org
O1: Before instruction, the participants completed the PCTS pre-test.
O2: Before instruction, the participants completed the IMMS pre-test.
O3: Before instruction, the participants completed the ASCOPL pre-test.
O4: After instruction, the participants completed the PCTS post-test.
O5: After instruction, the participants completed the IMMS post-test.
O6: After instruction, the participants completed the ASCOPL post-test.

*Programming Computational Thinking Scale (PCTS)*

The Programming Computational Thinking Scale developed by Wu Pei Zhen is employed in this research. (P. Z. Wu, 2021), designed for integration into a programming curriculum, it satisfies the criteria of this research. According to the study by Zhong et al. (Zhong et al., 2016), the MIT's three-dimensional computational thinking framework and this scale's design are closely related, including three dimensions. The questionnaire comprises 21 questions, with 8 covering Computational Concepts, 8 covering Computational Practices, and 5 covering Computational Perspectives. The scale is reliably measured, with a total Cronbach Alpha of .927.

*Instructional Materials Motivation Survey (IMMS)*

The IMMS was employed in this research, designed by Keller (Keller, 2010), to evaluate the motivation of students to learn programming. The IMMS consists of 36 questions distributed across four dimensions: Attention is assessed through 12 questions, Relevance through 9 questions, Confidence through 9 questions, and Satisfaction through another set of 6 questions. The IMMS is known for its good internal reliability and consistency in various educational contexts. The overall Cronbach Alpha is 0.924, indicating a substantial degree of reliability.

*Attitude Scale of Computer Programming Learning (ASCOPL)*

We employed Özgen Korkmaz's ASCOPL (Korkmaz & Altun, 2014), a tool known for its reliable and valid measurements. The scale effectively assessed the students' attitudes towards programming education. Comprising 20 questions, the scale is structured around three dimensions: 9 queries regarding Willingness, 6 for Negativity, and 5 for Necessity. The high level of reliability is confirmed by the scale's total Cronbach Alpha of .907.

**Programming Achievement on Code.org**

In this study, the evaluation of student achievement on Code.org was conducted through the Programming Achievement Test, which refers to the number of levels and stages completed by students on the website. This method allowed researchers to effectively monitor and quantify students' progress, generating summary tables that provided deep insights into the students' engagement and achievements within the curriculum. By focusing on completing levels, this approach provided a direct measure of the students' coding skills and problem-solving capabilities as fostered by the exercises on Code.org. It also served as a clear indicator of their overall academic success in computational thinking and programming.

*Data Analysis*

The instrument used for data analysis was the Statistical Package for Social Sciences (SPSS). Initially, a normality test was conducted. The findings indicated a normal distribution of the data. Subsequently, paired samples t-tests were employed to examine potential differences in pre-test and post-test scores for computational thinking, learning motivation, and learning attitudes. Academic achievements were compared with the Code.org website completion progress of individual students. Lastly, we used the Pearson correlation test to investigate relationships within the different scales and learning achievements in Code.org.

**Results**

*The Paired Sample T-test Result of PCTS*

To assess whether a notable difference existed between pre-test and post-test PCTS scores, a paired samples t-test was executed. The outcomes are presented in Table 2.

Table 2. PCTS paired samples t-test results

| Component | Stage | N | Mean | Sd | Se | t | p |
|---|---|---|---|---|---|---|---|
| Computational Concepts | Pre-test | 20 | 3.31 | 0.67 | 0.15 | -6.64 | 0.000 |
|  | Post-test | 20 | 3.96 | 0.52 | 0.12 |  |  |
| Computational Practices | Pre-test | 20 | 3.39 | 0.61 | 0.14 | -4.44 | 0.000 |
|  | Post-test | 20 | 3.91 | 0.55 | 0.12 |  |  |
| Computational Perspectives | Pre-test | 20 | 3.42 | 0.64 | 0.14 | -2.44 | 0.025 |
|  | Post-test | 20 | 3.70 | 0.59 | 0.13 |  |  |
| Total | Pre-test | 20 | 3.37 | 0.59 | 0.13 | -5.77 | 0.000 |
|  | Post-test | 20 | 3.88 | 0.51 | 0.11 |  |  |

In the dimension of Computational Concepts, there was a significant enhancement from the pre-test, with a mean score of 3.31 (SD = 0.67), to the post-test, where the mean score rose to 3.96 (SD = 0.52). This improvement was statistically significant, as the t-value of -6.64 with a p-value of less than 0.05 indicated a meaningful increase in students' understanding of computational concepts.

Similarly, the Computational Practices dimension showed a significant rise in the post-test mean score of 3.91 (SD = 0.55), compared to the pre-test mean score of 3.39 (SD = 0.61). The t-value of -4.44 with a p-value of less than 0.05 suggested that the instructional approach positively impacted students' computational practices.

The Computational Perspectives dimension, although less pronounced, still showed a significant improvement. The pre-test mean score of 3.42 (SD = 0.64) improved to 3.70 (SD = 0.59) in the post-test, with a t-value of -2.44 and a p-value less than 0.05, indicating a significant but modest enhancement in students' computational perspectives.

Overall, the total scores also reflected significant enhancement, with the pre-test score averaging 3.37 (SD = 0.59) and the post-test score at 3.88 (SD = 0.51). The t-value of -5.77, along with a p-value of less than 0.05, further substantiated the effectiveness of the intervention across all measured dimensions.

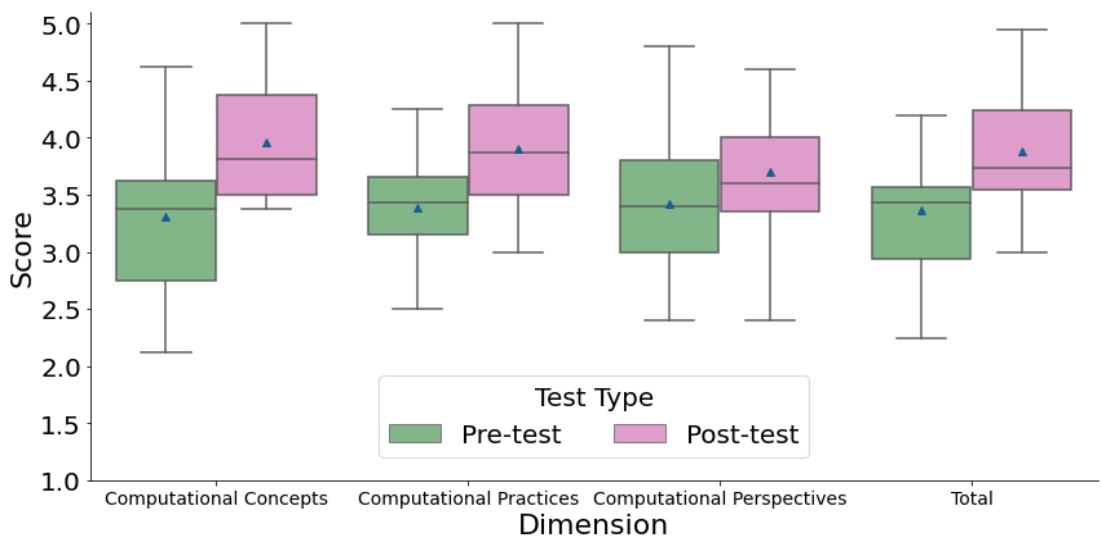

Figure 4. Boxplot for PCTS pre-test and post-test

The boxplot in Figure 4 visually compared pre-test and post-test scores across various dimensions of computational thinking abilities. The median scores for Computational Concepts, Computational Practices, Computational Perspectives, and the overall scores exhibited notable improvements from pre-test to post-test. This is depicted by the central triangle (the median) being higher in the post-test for these categories.

Specifically, the total scores and Computational Concepts scores showed a considerable increase in the median, alongside a more uniform distribution as indicated by the tighter interquartile range (IQR) in the post-test results, implying a more consistent performance among the participants following the educational intervention.

In the dimension of Computational Practices, while the median score did improve in the post-test, the IQR slightly widened, indicating a more significant variation in participant scores after the intervention. This suggests that while overall proficiency increased, the variance in individual performance also became more pronounced.

For Computational Perspectives, a slight increase in the median score was observed from pre-test to post-test. However, the IQR remained relatively unchanged, indicating that the spread of scores did not significantly alter. Hence, the effect of the intervention on this dimension was less substantial than that of the other areas.

The smallest and largest observed scores (depicted by the whiskers of the boxplot) across all dimensions show improvements in both the lower and upper bounds from pre-test to post-test, with the post-test scores reaching the maximum score of 5 in several instances.

Overall, the boxplot underscores a positive score trend post-intervention, particularly in Computational Concepts and Practices. While present, the enhancement in Computational Perspectives was not as marked as in the other dimensions.

These results collectively indicated that the educational program implemented with Code.org between the pre-test and post-test was highly influential in enhancing students' computational thinking skills across all measured dimensions. Factors contributing to this success may have included an engaging curriculum, effective teaching methodologies, or the integration of practical exercises that enhanced the understanding and application of computational concepts, practices, and perspectives. Additionally, the structured approach of Code.org programming could have played a pivotal role in promoting computational thinking in students.

The Computational Concepts dimension could have been enhanced by the clear and concise Code.org programming lessons that demystified complex concepts, making them more accessible for students. The Computational Practices dimension has been improved due to the hands-on coding exercises that Code.org provides, allowing students to apply theoretical knowledge in a practical context. As for the Computational Perspectives dimension, the peer interaction facilitated by Code.org could have broadened students' views on significantly improving computational thinking.

*The Paired Sample T-test result of IMMS*

The paired samples t-test was conducted to ascertain the presence of a significant difference between pre-test and post-test IMMS scores, as shown in Table 3.

Table 3. IMMS paired samples t-test results

| Component | Stage | N | Mean | Sd | Se | t | p |
|---|---|---|---|---|---|---|---|
| Attention | Pre-test | 20 | 3.72 | 0.39 | 0.09 | -5.24 | 0.000 |
| | Post-test | 20 | 4.17 | 0.58 | 0.13 | | |
| Relevance | Pre-test | 20 | 3.74 | 0.48 | 0.11 | -6.17 | 0.000 |
| | Post-test | 20 | 4.19 | 0.50 | 0.11 | | |
| Confidence | Pre-test | 20 | 3.52 | 0.37 | 0.08 | -4.65 | 0.000 |
| | Post-test | 20 | 3.93 | 0.36 | 0.08 | | |
| Satisfaction | Pre-test | 20 | 3.87 | 0.38 | 0.09 | -4.53 | 0.000 |
| | Post-test | 20 | 4.36 | 0.47 | 0.10 | | |
| Total | Pre-test | 20 | 3.71 | 0.36 | 0.08 | -6.82 | 0.000 |
| | Post-test | 20 | 4.17 | 0.42 | 0.09 | | |

Starting with the Attention dimension, there was a noteworthy improvement from the pre-test, which had a mean score of 3.72 (SD = 0.39), to the post-test, where the mean score increased to 4.17 (SD = 0.58). The t-value of -5.24 and a p-value less than 0.05 significantly enhanced how the instructional material captured and maintained the learners' Attention.

In the Relevance dimension, the post-test mean score rose to 4.19 (SD = 0.50) from a pre-test mean score of 3.74 (SD = 0.48). The t-value of -6.17 with a p-value less than 0.05 suggested a substantial increase in the perceived Relevance of the material to the learners' needs and interests.

For the Confidence dimension, there was a significant increase in the post-test score, with a mean of 3.93 (SD = 0.36), compared to the pre-test score of 3.52 (SD = 0.37). The t-value of -4.65 and a p-value less than 0.05 indicated that the material effectively boosted learners' Confidence in their ability to succeed.

The Satisfaction dimension also showed significant growth, with the pre-test mean score of 3.87 (SD = 0.38) rising to a post-test score of 4.36 (SD = 0.47). The t-value of -4.53 and a p-value less than 0.05 demonstrated a marked increase in learner satisfaction with the instructional material.

Overall, the total IMMS scores reflected a significant enhancement, with the average pre-test score of 3.71 (SD = 0.36) increasing to a post-test score of 4.17 (SD = 0.42). The t-value of -6.82 and a p-value less than 0.05 further validated the effectiveness of the intervention across all measured dimensions.

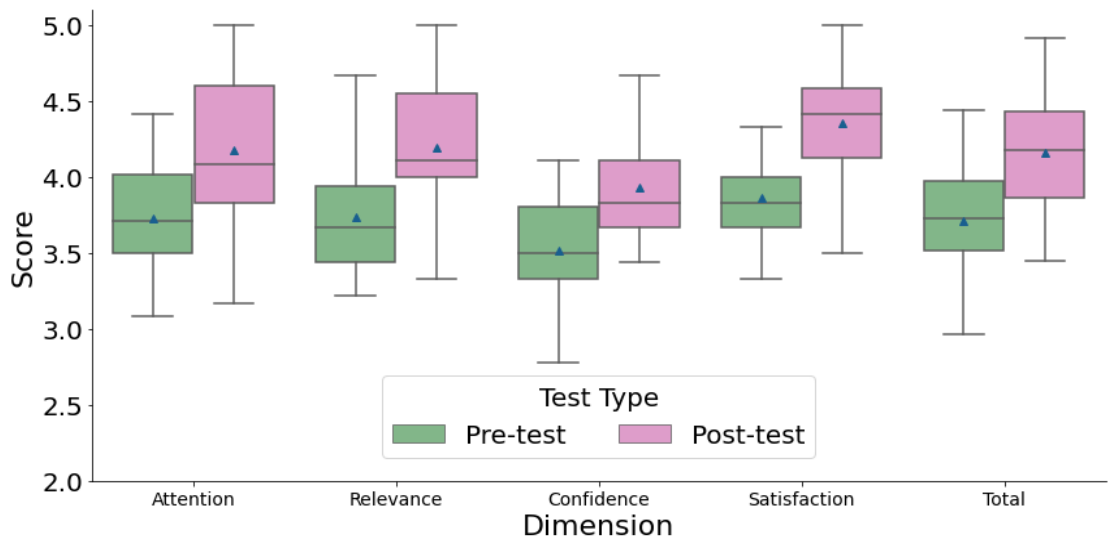

Figure 5. Boxplot for IMMS pre-test and post-test

The boxplot visualization in Figure 5 indicated notable differences between the pre-test and post-test scores across various dimensions of student motivation. Focusing on the Attention dimension, the post-test scores exhibited an uplift in the median, indicating increased attention post-intervention. The range of scores, as suggested by the wider IQR, also expanded, reflecting a greater diversity in students' attention levels after the intervention.

In the Confidence dimension, there was a clear upward shift in the post-test median score, with the spread of scores also extending further. This suggests that students' confidence increased on average, but there was variability in how much confidence each student gained.

The Relevance dimension also showed a higher median in the post-test scores, with a more concentrated clustering of scores around the median, indicating a more uniform agreement on the material's relevance among the students after the intervention.

For Satisfaction, the post-test scores again showed a higher median than the pre-test, with a greater spread of scores, denoting that while satisfaction generally improved, the degree of satisfaction varied among the students.

Overall, the Total dimension captured the aggregated effect of the intervention, revealing a significant rise in the median score and a moderate increase in the IQR in the post-test results, which aligned with the substantial improvements noted.

The visual data from these boxplots succinctly represented the positive impact of the educational intervention on all aspects of student motivation. There was an apparent increase in median values and variations in score distributions, highlighting the intervention's varying influence on the students' motivational experiences.

These results collectively suggested that the instructional material or program implemented between the pre-test and post-test was highly influential in enhancing the motivational aspects of learning across all measured dimensions. The significant improvements across Attention, Relevance, Confidence, and Satisfaction dimensions indicate that the material successfully engaged and motivated the learners. Contributing factors to this success could include the Relevance and applicability of the content, interactive and engaging instructional strategies, and the alignment of the material with learners' needs and goals.

Concerning the four dimensions of learning motivation - Attention, Relevance, Confidence, and Satisfaction- the instructional material of Code.org addressed each aspect effectively. The Attention dimension was likely enhanced through interactive programming tasks that stimulated interest and curiosity. The Relevance dimension was probably promoted by providing real-world applications of coding concepts, making the learning material meaningful and directly applicable to the learners' lives. The Confidence dimension was likely boosted by the program's scaffolded learning approach, which progressively built learners' skills and Confidence in coding. Lastly, the Satisfaction dimension was probably increased by the sense of achievement the learners experienced upon completing the programming tasks, as well as the overall engaging and supportive learning environment fostered by the program.

*The Paired Sample T-test result of ASCOPL*

The study employed a paired samples t-test to evaluate whether notable changes occurred in the ASCOPL between the pre-test and post-test. Outcomes are presented in Table 4.

Table 4. ASCOPL paired samples t-test results

| Component | Stage | N | Mean | Sd | Se | t | p |
|---|---|---|---|---|---|---|---|
| Willingness | Pre-test | 20 | 3.62 | 0.55 | 0.12 | -6.28 | 0.000 |
| | Post-test | 20 | 4.03 | 0.51 | 0.11 | | |
| Negativity | Pre-test | 20 | 3.80 | 0.62 | 0.14 | -3.61 | 0.002 |
| | Post-test | 20 | 4.18 | 0.83 | 0.19 | | |
| Necessity | Pre-test | 20 | 3.82 | 0.76 | 0.17 | -2.27 | 0.035 |
| | Post-test | 20 | 4.10 | 0.75 | 0.17 | | |
| Total | Pre-test | 20 | 3.73 | 0.52 | 0.12 | -5.71 | 0.000 |
| | Post-test | 20 | 4.09 | 0.58 | 0.13 | | |

An appreciable enhancement in the Willingness dimension was noted in contrast to the pre-test, with a mean score of 3.62 (SD = 0.55), to the post-test, where the mean score rose to 4.03 (SD = 0.51). This rise was statistically significant, as evidenced by the t-test result of -6.28 and a p-value less than 0.05, indicating a substantial increase in participants' willingness levels.

Similarly, in the Negativity dimension, the post-test mean score increased to 4.18 (SD = 0.83) from a pre-test mean score of 3.80 (SD = 0.62). The t-value of -3.61 with a p-value less than 0.05 suggested that the intervention positively impacted reducing negativity among participants.

The Necessity dimension showed a less pronounced yet significant improvement. The pre-test mean score of 3.82 (SD = 0.76) improved to 4.10 (SD = 0.75) in the post-test. The t-value of -2.27 and a p-value less than 0.05 indicated a significant enhancement in the perceived necessity of whatever construct or behavior ASCOPL measures.

Overall, the total scores across all dimensions also reflected a significant enhancement. The pre-test score averaged 3.73 (SD = 0.52) and improved to 4.09 (SD = 0.58) in the post-test. The t-value of -5.71 and a p-value less than 0.05 provided strong evidence of the intervention's effectiveness between the pre-test and post-test periods.

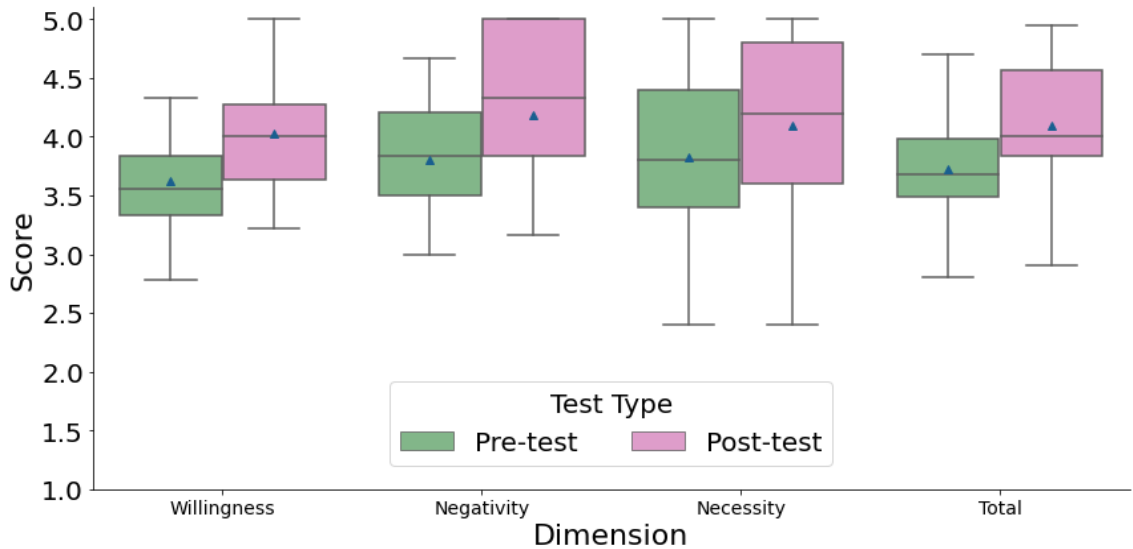

Figure 6. Boxplot for ASCOPL pre-test and post-test

The boxplots in Figure 6 indicated an overall improvement from the pre-test to the post-test across all dimensions measured.

For the Necessity dimension, the post-test median score was notably higher than the pre-test median, suggesting an increased recognition of the necessity of the subject matter among the participants. The broader range in the post-test, as evidenced by a larger IQR, indicates a greater diversity in the participants' responses after the intervention.

The Negativity dimension showed a noticeable increase in the median score from the pre-test to the post-test. The distribution of scores post-test was wider, with a higher maximum, implying a decrease in negative sentiments towards the subject after the intervention.

The Willingness dimension increased both the median score and the spread of scores post-test. This suggests that participants were generally more willing to engage with the subject matter after the intervention, with some experiencing a significant increase in willingness.

Regarding the Total scores, the post-test showed a modest rise in the median along with a slight increase in the IQR. The minimum and maximum scores also covered a wider range post-test, indicating an overall improvement in the total assessment of the intervention's impact.

These results collectively suggested that the intervention or educational program associated with ASCOPL greatly enhanced the targeted attributes or skills across all measured dimensions. The significant improvements implied that the methodologies successfully improved specific attributes measured by ASCOPL and fostered an overall increase in the positive aspects of these dimensions.

In line with Code.org's focus on fostering a positive learning attitude, there is an association between the dimensions of Willingness, Negativity, and Necessity and the principles of perseverance, resilience, and recognition of the importance of the subject matter, respectively. The significant improvement in these areas may be attributed to Code.org's comprehensive curriculum emphasizing problem-solving, creativity, and collaboration. The interactive and engaging nature of the platform may have also contributed to the increased willingness and reduced negativity towards learning. The improved necessity scores may reflect an increased understanding of the importance and Relevance of coding skills in today's digital age.

### *Student Achievement*

The teacher dashboard on Code.org was an essential tool for monitoring student progress. It provided a range of features for managing and assessing student activities in specific areas, including reviewing and grading their work. Information regarding the stages and levels completed was gathered to evaluate student performance. This analysis revealed that students' achievements during the three-week course on Code.org significantly exceeded their performance in traditional classroom settings. Students exerted considerable dedication to grasp computational thinking concepts and apply them to various programming tasks.

Table 5. Completion statistics by lesson

| Lesson | Programming Concepts | Total Level | Completed Percentage |
|---|---|---|---|
| 1. Graph Paper Programming | Unplugged Activity | 2 | 100% |
| 2. Real-Life Algorithms: Paper Airplanes | Unplugged Activity | 2 | 100% |
| 3. Maze1 - Sequence Basics | Sequential Logic | 11 | 95% |
| 4. Artist1: Introduction to Sequences | Sequential Logic | 12 | 87% |
| 5. Getting Loopy | Unplugged Activity | 1 | 100% |
| 6. Maze2: Loop Concepts | Loop Programming | 14 | 100% |
| 7. Artist2: Exploring Loops | Loop Programming | 16 | 98% |
| 8. Bee1: Loop Challenges | Loop Programming | 14 | 98% |
| 9. Relay Programming | Unplugged Activity | 2 | 100% |
| 10. Bee2: The Art of Debugging | Debugging Skills Development | 11 | 99% |
| 11. Artist3: Debugging Techniques | Debugging Skills Development | 12 | 81% |
| 12. Conditional Logic with Cards | Unplugged Activity | 1 | 100% |
| 13. Bee3: Mastering Conditionals | Conditional Programming | 15 | 97% |
| 14. Binary Bracelets | Unplugged Activity | 1 | 100% |
| 15. The Big Event Planning | Unplugged Activity | 1 | 100% |
| 16. Flappy: Events and Parallelism | Event and Parallel Programming | 10 | 89% |
| 17. Create a Story in Play Lab | Event and Parallel Programming | 11 | 75% |
| 18. Your Digital Footprint Matters | Unplugged Activity | 1 | 100% |
| 19. Artist4: Nested Loops Challenge | Advanced Nested Loop | 13 | 33% |

Based on Table 5, all students successfully completed the unplugged activities, including lessons 1, 2, 5, 9, 12, 14, 15, and 18. High completion rates were observed in lessons involving loop programming and conditional programming, with Maze2 (lesson 6), Artist2 (lesson 7), Bee1 (lesson 8), and Bee3 (lesson 13) achieving rates of 100%, 98%, 98%, and 97%, respectively. Sequential logic and debugging skills lessons also showed strong performance, with Maze1 (lesson 3) at a 95% completion rate, and both Bee2 (lesson 10) and Artist3 (lesson 11) above 80%. Artist 1: Introduction to Sequences (lesson 4) had an 87% completion rate. The completion rates for event and parallel programming in Flappy (lesson 16) and Play Lab: Create a Story (lesson 17) were 89% and 75%, respectively. However, Artist4 (lesson 19), focusing on advanced nested loops, had a much lower completion rate at 33%, highlighting its complexity and the potential need for more effective teaching strategies or additional learning time for these advanced concepts.

Further analysis, as shown in Table 6, provided individual completion statistics, revealing a wide range of completion percentages among participants, from 60% to 100%. This variance suggested that while most students thrived in the Code.org environment, a smaller group found certain materials more challenging. Factors contributing to this disparity included the complexity of stages like Artist4 and students' varying learning paces and prior knowledge.

Nonetheless, the overall trend observed in Table 6 was positive. Most students demonstrated high levels of engagement and success, with many achieving near-perfect completion rates. This was particularly notable considering the diversity of tasks, which ranged from basic unplugged activities to more complex programming challenges.

Table 6. Completion statistics by participant

| Participant | Completed Level | Completed Percentage |
|---|---|---|
| P01 | 147 | 98.0% |
| P02 | 125 | 83.3% |
| P03 | 149 | 99.3% |
| P04 | 126 | 84.0% |
| P05 | 132 | 88.0% |
| P06 | 133 | 88.7% |
| P07 | 150 | 100.0% |
| P08 | 108 | 72.0% |
| P09 | 138 | 92.0% |
| P10 | 131 | 87.3% |
| P11 | 133 | 88.7% |
| P12 | 137 | 91.3% |
| P13 | 90 | 60.0% |
| P14 | 149 | 99.3% |
| P15 | 136 | 90.7% |
| P16 | 125 | 83.3% |
| P17 | 141 | 94.0% |
| P18 | 140 | 93.3% |
| P19 | 125 | 83.3% |
| P20 | 127 | 84.7% |

Moreover, in preparation for future analysis with ANOVA, we calculated the quartiles to categorize the PAT scores into three segments: high, medium, and low. The first quartile (Q1) was determined to be 83.50, the median (Q2) was 88.67, and the third quartile (Q3) was 93.83.

Accordingly, the grouping criteria were established as follows: students scoring less than or equal to 83.50 were placed in the Low group, those with scores above 83.50 but less than or equal to 93.83 points in the Medium group, and students with scores above 93.83 points in the High group.

Table 7. Descriptive statistics results of PAT

| Group | N | Median | Mean | Sd |
|---|---|---|---|---|
| Low | 10 | 83.50 | 76.40 | 9.80 |
| Medium | 20 | 88.67 | 88.87 | 2.97 |
| High | 10 | 93.83 | 98.13 | 2.28 |

The distribution of students and PAT scores across three groups is detailed in Table 7. The Low group consisted of 10 students, with a mean of 76.40, and a standard deviation (SD) of 9.80. The Medium group includes 20 students, with a mean of 88.87, and an SD of 2.97. Finally, the High group comprises 10 students, with a mean of 98.13, and an SD of 2.28.

Overall, the combined analysis offered a comprehensive view of the learning outcomes achieved through Code.org. It underscored the platform's effectiveness in imparting critical computational skills and highlighted areas where further support was required, especially for students facing difficulties with advanced concepts. While the program was largely successful, the data indicated the need for tailored approaches to meet the diverse needs of individual learners.

*The correlation between PCTS, IMMS, ASCOPL, and achievement in Code.org*

We aimed to establish correlations between PCTS, IMMS, ASCOPL, and achievement in Code.org, employing the Pearson correlation test, the results of which are detailed in Figure 7.

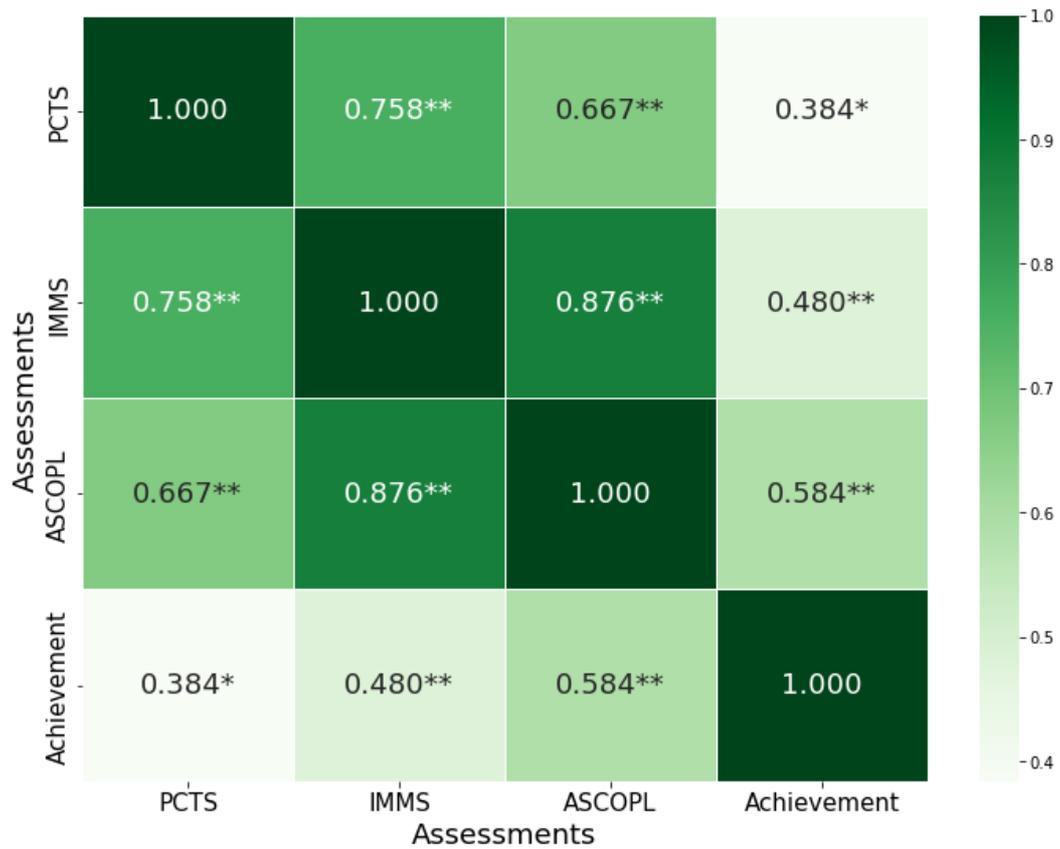

** indicates significance at the 0.01 level, * at the 0.05 level.

Figure 7. Correlation between PCTS, IMMS, ASCOPL, and Achievement

The analysis was oriented from highest to lowest correlation to elucidate the relationships between PCTS, IMMS, ASCOPL, and Achievement in Code.org.

*1)    Correlation between ASCOPL and Achievement*

The strongest correlation was found between ASCOPL and Achievement (.584**). This indicates the performance outcomes on Code.org were significantly tied to students' attitudes toward learning computer programming. A positive learning attitude greatly enhanced students' ability to understand programming concepts and succeed in Code.org challenges.

*2)    Correlation between IMMS and Achievement*

The following significant correlation was observed between IMMS and Achievement

(.480**). This suggests that the motivation derived from instructional materials was crucial in influencing student performance. The effectiveness and engagement of the instructional materials on Code.org substantially impacted students' motivation to learn, thereby affecting their achievement.

*3)     Correlation between PCTS and Achievement*

The correlation between PCTS and Achievement was somewhat lower (.384*), indicating that while computational thinking skills and programming knowledge were critical, their impact on Code.org performance was less pronounced than ASCOPL and IMMS. This finding implied that although computational thinking was vital for programming, the attitude toward learning and the motivational quality of instructional materials played more critical roles in students' performance.

*4)     Correlation between PCTS, IMMS, ASCOPL, and Achievement*

The PCTS, IMMS, and ASCOPL interrelationships indicated a complex, interconnected dynamic. These three components - computational thinking skills, the motivational quality of instructional materials, and students' attitudes toward programming - were deeply intertwined.

The strong correlations suggested that these elements influenced and reinforced each other. For example, high-quality, engaging instructional materials (IMMS) motivated students and positively shaped their attitudes toward programming (ASCOPL). A positive attitude enhanced their engagement with and understanding of computational thinking concepts (PCTS). Similarly, as students developed more vital computational thinking skills, this proficiency positively influenced their attitudes toward programming. It made them more receptive to the instructional materials, creating a reinforcing cycle of motivation, attitude, and skill enhancement.

This interplay suggested that a practical approach in programming education should have simultaneously addressed these three areas. Enhancing computational thinking skills was not just about direct instruction but also about creating motivating instructional materials and fostering positive attitudes toward programming. The synergy among PCTS, IMMS, and ASCOPL highlighted the importance of an integrated educational strategy that considered how these elements interacted to influence students' learning outcomes in programming.

*ANOVA Analysis for PCTS and PAT*

The ANOVA results presented in Table 8 were used to determine if there were any statistically significant differences between the means of the different groups for PCTS and PAT.

Table 8. ANOVA results for PCTS and PAT

| Dimension | Groups | SS | Df | MS | F | p |
|---|---|---|---|---|---|---|
| Computational Concepts | Between groups | 4.93 | 2 | 2.46 | 7.064 | 0.003* |
| | Within groups | 12.90 | 37 | 0.35 | | |
| | Total | 17.82 | 39 | -- | | |
| Computational Practices | Between groups | 2.19 | 2 | 1.09 | 3.056 | 0.059 |
| | Within groups | 13.25 | 37 | 0.36 | | |
| | Total | 15.43 | 39 | -- | | |
| Computational Perspectives | Between groups | 3.74 | 2 | 1.87 | 6.063 | 0.005* |
| | Within groups | 11.40 | 37 | 0.31 | | |
| | Total | 15.14 | 39 | -- | | |
| Overall | Between groups | 3.43 | 2 | 1.72 | 5.928 | 0.006* |
| | Within groups | 10.72 | 37 | 0.29 | | |
| | Total | 14.15 | 39 | -- | | |

* denoted that $p < 0.05$

Computational Concepts: The ANOVA results indicated a statistically significant difference between groups in the domain of Computational Concepts, with an F-ratio of 7.064 and a p-value of 0.003. This suggested significant differences in the understanding or applying computational concepts across the groups studied.

Computational Practices: The results showed an F-ratio of 3.056 with a p-value of 0.059. This indicated that, although there was some variance in computational practices among the groups, the differences were not statistically significant at the 0.05 level, suggesting more similarity than difference in this domain across the groups.

Computational Perspectives: This dimension showed a statistically significant difference between groups, with an F-ratio of 6.063 and a p-value of 0.005. This finding implied that students' perspectives toward programming varied significantly across different groups, highlighting differences in how these groups engaged with or perceived computational thinking and practices.

The overall PCTS score showed a significant difference between the groups, as indicated by an F-ratio of 5.928 and a p-value of 0.006. This significant difference suggested that, across all domains considered, there were meaningful variations in computational thinking abilities among the groups, pointing to the impact of differing educational or experiential backgrounds on students' computational competencies and perspectives.

Following the ANOVA analysis, in-depth pairwise comparisons were conducted to detail the differences between these groups. These comparisons are outlined in Tables 9 to 11 using the Bonferroni method for a rigorous evaluation.

Table 9's pairwise comparisons within the Computational Concepts dimension revealed meaningful differences. The High group's mean score of 4.24 significantly exceeded both the Low group's score (mean = 3.36, difference of 0.88, $p < 0.05$) and the Medium group's score (mean = 3.47, difference of 0.77, $p < 0.05$). This highlighted a graded relationship between PAT scores and understanding of computational concepts, suggesting that higher achievement in programming is linked to a deeper grasp of these core concepts across groups.

Table 9. Pairwise comparisons of computational concepts among different groups for PCTS and PAT

| Groups | Mean | Low | Medium | High |
|---|---|---|---|---|
| Low | 3.36 | -- | -0.11 | -0.88* |
| Medium | 3.47 | 0.11 | -- | -0.77* |
| High | 4.24 | 0.88* | 0.77* | -- |

* denoted that p < 0.05

Similarly, from Table 10, the increase in the Computational Perspectives dimension from the Low group (mean = 3.48) to the High group (mean = 4.08) by 0.60 points (p < 0.05) suggested that students with higher PAT scores also held more positive attitudes towards programming.

Table 10. Pairwise comparisons of computational perspectives among different groups for PCTS and PAT

| Groups | Mean | Low | Medium | High |
|---|---|---|---|---|
| Low | 3.48 | -- | 0.14 | -0.60 |
| Medium | 3.34 | -0.14 | -- | -0.74* |
| High | 4.08 | 0.60 | 0.74* | -- |

* denoted that p < 0.05

The overall scores in Table 11, derived from the PCTS for assessing computational thinking abilities, show that the High group, with an average score of 4.13, markedly outperformed the Low group's average of 3.43 by 0.70 points (p < 0.05), and the Medium group's average of 3.46 by 0.67 points (p < 0.05). This distinction underscores the significant impact of comprehensive computational thinking skills on students' programming achievements.

Table 11. Pairwise comparisons of overall scores among different groups for PCTS and PAT

| Groups | Mean | Low | Medium | High |
|---|---|---|---|---|
| Low | 3.43 | -- | -0.03 | -0.70* |
| Medium | 3.46 | 0.03 | -- | -0.67* |
| High | 4.13 | 0.70* | 0.67* | -- |

* denoted that p < 0.05

In summary, the ANOVA analysis revealed statistically significant differences, especially in Computational Concepts and Computational Perspectives. Although no statistically significant variation was observed in Computational Practices among the groups, the results offer critical insights for educational strategies. Specifically, they highlight the importance of prioritizing enhancing computational concepts and perspectives to improve students' programming achievements. This recommendation is based on the strong positive correlation between these domains and programming achievement across various groups. Therefore, focusing on these areas is crucial for educational interventions designed to bridge the gap in computational thinking skills among diverse groups.

*ANOVA Analysis for IMMS and PAT*

The ANOVA analysis presented in Table 12 aimed to determine if there were any statistically significant differences between the means of different groups for IMMS and PAT.

Table 12. ANOVA results for IMMS and PAT

| Dimension | Groups | SS | Df | MS | F | p |
| --- | --- | --- | --- | --- | --- | --- |
| Attention | Between groups | 2.08 | 2 | 1.04 | 4.162 | 0.023* |
| | Within groups | 9.23 | 37 | 0.25 | | |
| | Total | 11.30 | 39 | -- | | |
| Relevance | Between groups | 2.63 | 2 | 1.32 | 5.746 | 0.007* |
| | Within groups | 8.48 | 37 | 0.23 | | |
| | Total | 11.12 | 39 | -- | | |
| Confidence | Between groups | 1.34 | 2 | 0.67 | 4.527 | 0.017* |
| | Within groups | 5.46 | 37 | 0.15 | | |
| | Total | 6.79 | 39 | -- | | |
| Satisfaction | Between groups | 0.79 | 2 | 0.39 | 1.705 | 0.196 |
| | Within groups | 8.57 | 37 | 0.23 | | |
| | Total | 9.35 | 39 | -- | | |
| Overall | Between groups | 1.60 | 2 | 0.80 | 4.700 | 0.015* |
| | Within groups | 6.28 | 37 | 0.17 | | |
| | Total | 7.87 | 39 | -- | | |

* denoted that $p < 0.05$

Attention: The ANOVA results indicated a statistically significant difference between groups in the domain of Attention, with an F-ratio of 4.162 and a p-value of 0.023. This suggested meaningful differences in how groups focused on and engaged with programming materials, underscoring the variability in attentional engagement across the studied groups.

Relevance: In the Relevance dimension, the analysis revealed a statistically significant difference between groups, with an F-ratio of 5.746 and a p-value of 0.007. This finding implied that students' perceptions of the relevance of programming materials to their learning and future goals significantly differed across groups, highlighting the importance of aligning instructional materials with students' interests and career aspirations.

Confidence: The Confidence dimension showed a statistically significant difference between groups, with an F-ratio of 4.527 and a p-value of 0.017. This indicates that students' confidence in their programming abilities varies significantly across different groups, pointing to the impact of self-efficacy beliefs on programming learning outcomes.

Satisfaction: The results showed an F-ratio of 1.705 with a p-value of 0.196 for the dimension of Satisfaction. This indicates that, although there is some variance in satisfaction among the groups, the differences were not statistically significant at the 0.05 level. This suggests more similarity than difference in satisfaction with programming learning experiences across the groups.

The overall IMMS score indicated a statistically significant difference between the groups, as shown by an F-ratio of 4.700 and a p-value of 0.015. This overall significant difference suggests that, across all dimensions considered, there are meaningful variations in the motivational aspects of instructional materials among the groups. This points to the impact of differing educational or experiential backgrounds on students' engagement and motivation towards programming education.

Following the ANOVA analysis, in-depth pairwise comparisons were conducted to detail the differences between these groups. These comparisons are outlined in Tables 14 to 17 using the Bonferroni method for a rigorous evaluation.

The detailed analysis of pairwise comparisons for the Attention dimension, as presented in Table 13, showed that the High group, with a mean score of 4.34, significantly surpassed both the Low group, with a mean score of 3.77 (difference of 0.57, $p < 0.05$), and the Medium group, with a mean score of 3.84 (difference of 0.50, $p < 0.05$). This indicates that students with higher PAT scores are more focused and engaged with the programming materials, suggesting that attentional engagement is linked to higher achievement in programming.

Table 13. Pairwise comparisons of Attention among different groups for IMMS and PAT

| Groups | Mean | Low | Medium | High |
| --- | --- | --- | --- | --- |
| Low | 3.77 | -- | -0.07 | -0.57* |
| Medium | 3.84 | 0.07 | -- | -0.50* |
| High | 4.34 | 0.57* | 0.50* | -- |

* denoted that $p < 0.05$

In Table 14, pairwise comparisons within the Relevance dimension revealed that the High group's mean score of 4.41 markedly exceeded that of the Low and Medium groups, both with mean scores of 3.82 (difference of 0.59, $p < 0.05$). This underscored the importance of the perceived relevance of programming materials in students' learning. It indicated that students who saw higher relevance in their materials tended to achieve better in programming.

Table 14. Pairwise comparisons of Relevance among different groups for IMMS and PAT

| Groups | Mean | Low | Medium | High |
| --- | --- | --- | --- | --- |
| Low | 3.82 | -- | 0.01 | -0.59* |
| Medium | 3.82 | -0.01 | -- | -0.59* |
| High | 4.41 | 0.59* | 0.59* | -- |

* denoted that $p < 0.05$

Table 15's comparisons showed significant differences in the Confidence dimension. The High group, with a mean score of 4.03, significantly outperformed the Low group, with a mean of 3.56 (difference of 0.48, $p < 0.05$), and the Medium group, with a mean of 3.66 (difference of 0.38, $p < 0.05$). These results highlighted the critical

role of confidence in programming abilities, suggesting that enhancing students' self-efficacy may lead to better programming outcomes.

Table 15. Pairwise comparisons of Confidence among different groups for IMMS and PAT

| Groups | Mean | Low | Medium | High |
|---|---|---|---|---|
| Low | 3.56 | -- | -0.10 | -0.48* |
| Medium | 3.66 | 0.10 | -- | -0.38* |
| High | 4.03 | 0.48* | 0.38* | -- |

* denoted that $p < 0.05$

Finally, the comparisons for the Overall IMMS score, detailed in Table 16, showed that the High group, with a mean score of 4.28, significantly exceeded the Low group, with a mean of 3.81 (difference of 0.48, $p < 0.05$), and the Medium group, with a mean of 3.83 (difference of 0.45, $p < 0.05$). This indicated that higher PAT scores are associated with a more positive overall perception and motivation toward programming education.

Table 16. Pairwise comparisons of overall scores among different groups for IMMS and PAT

| Groups | Mean | Low | Medium | High |
|---|---|---|---|---|
| Low | 3.81 | -- | -0.02 | -0.48* |
| Medium | 3.83 | 0.02 | -- | -0.45* |
| High | 4.28 | 0.48* | 0.45* | -- |

* denoted that $p < 0.05$

Following the pairwise comparisons from Tables 13 to 16, these updated analyses elucidated the nuanced relationship between various dimensions of the IMMS and students' achievement levels as measured by PAT scores. The findings suggested that not only individual components like Attention, Relevance, and Confidence played significant roles in programming achievement but also that a positive overall programming experience was a crucial predictor of success. These insights advocated for educational strategies that focused on enhancing specific aspects of programming instruction to improve engagement and self-efficacy and aimed to create a positive overall learning experience to boost academic outcomes in programming.

*ANOVA Analysis for ASCOPL and PAT*

The ANOVA analysis presented in Table 17 aimed to determine if there were any statistically significant differences between the means of different groups for ASCOPL and PAT.

Table 17. ANOVA results for ASCOPL and PAT

| Dimension | Groups | SS | Df | MS | F | p |
|---|---|---|---|---|---|---|
| Willingness | Between groups | 2.99 | 2 | 1.49 | 5.904 | 0.006* |
| | Within groups | 9.36 | 37 | 0.25 | | |
| | Total | 12.34 | 39 | -- | | |
| Negativity | Between groups | 4.52 | 2 | 2.26 | 4.851 | 0.013* |
| | Within groups | 17.25 | 37 | 0.47 | | |
| | Total | 21.78 | 39 | -- | | |
| Necessity | Between groups | 4.27 | 2 | 2.14 | 4.337 | 0.020* |
| | Within groups | 18.22 | 37 | 0.49 | | |
| | Total | 22.50 | 39 | -- | | |
| Overall | Between groups | 3.05 | 2 | 1.53 | 5.743 | 0.007* |
| | Within groups | 9.84 | 37 | 0.27 | | |
| | Total | 12.89 | 39 | -- | | |

* denoted that $p < 0.05$

Willingness: The ANOVA results revealed a statistically significant difference between groups, with an F-ratio of 5.904 and a p-value of 0.006. This indicated that the eagerness to engage with programming learning varied significantly across the groups, suggesting that students' willingness to learn programming could significantly influence their achievements in this field.

Negativity: In the dimension of Negativity, the analysis showed a significant difference between groups, evidenced by an F-ratio of 4.851 and a p-value of 0.013. This finding suggested that negative attitudes towards programming varied across different groups, significantly impacting programming learning outcomes. Reducing negative perceptions and increasing positive engagement with programming could enhance students' learning experiences and outcomes.

Necessity: The results demonstrated a statistically significant difference between groups, with an F-ratio of 4.337 and a p-value of 0.020. This underscored the importance of perceiving programming as necessary for one's education or career, suggesting that students who recognized the necessity of programming were more likely to achieve higher levels in this discipline.

The overall ASCOPL score represented a cumulative assessment of attitudes toward computer programming learning. It showed a significant difference between groups, as indicated by an F-ratio of 5.743 and a p-value of 0.007. This significant variance highlights the impact of comprehensive attitudes towards programming on students' achievements, pointing to the critical role of cultivating positive attitudes for enhancing programming learning outcomes.

Following the ANOVA analysis, which identified significant differences across PAT score groups in various PCTS dimensions, in-depth pairwise comparisons were conducted to detail the differences between these groups. Employing the Bonferroni method for a rigorous evaluation, these comparisons are outlined in Tables 19 to 22.

Table 18's pairwise comparisons within the Willingness dimension revealed that the High group, with an average score of 4.26, significantly outperformed both the Low group, with an average score of 3.87 (difference of 0.39, $p < 0.05$), and the Medium group, with an average score of 3.59 (difference of 0.67, $p < 0.05$). These findings suggested a positive correlation between students' willingness to engage with programming and their achievement levels, highlighting the importance of fostering a proactive engagement towards programming learning.

Table 18. Pairwise comparisons of Willingness among different groups for ASCOPL and PAT

| Groups | Mean | Low | Medium | High |
|---|---|---|---|---|
| Low | 3.87 | -- | 0.28 | -0.39 |
| Medium | 3.59 | -0.28 | -- | -0.67* |
| High | 4.26 | 0.39 | 0.67* | -- |

* denoted that $p < 0.05$

Table 19 focused on the Negativity dimension, showing that the High group, with an average score of 4.48, exhibited significantly lower levels of negative attitudes towards programming compared to the Low group, which had an average score of 3.53 (difference of 0.95, p < 0.05), and the Medium group, with an average score of 3.98 (difference of 0.50, p < 0.05). These results underline the critical role of reducing negative perceptions and enhancing positive engagement with programming to improve learning outcomes.

Table 19. Pairwise comparisons of Negativity among different groups for ASCOPL and PAT

| Groups | Mean | Low | Medium | High |
|---|---|---|---|---|
| Low | 3.53 | -- | -0.44 | -0.95* |
| Medium | 3.98 | 0.44 | -- | -0.51 |
| High | 4.48 | 0.95* | 0.51 | -- |

* denoted that p < 0.05

In the Necessity dimension, outlined in Table 20, the High group's average score of 4.50 was significantly higher than that of the Low group, with an average of 3.62 (difference of 0.88, p < 0.05), and the Medium group, with an average of 3.86 (difference of 0.64). This demonstrates the impact of perceiving programming as a necessary skill for education or career, suggesting that recognizing the importance of programming is associated with higher achievement levels.

Table 20. Pairwise comparisons of Necessity among different groups for ASCOPL and PAT

| Groups | Mean | Low | Medium | High |
|---|---|---|---|---|
| Low | 3.62 | -- | -0.24 | -0.88* |
| Medium | 3.86 | 0.24 | -- | -0.64 |
| High | 4.50 | 0.88* | 0.64 | -- |

* denoted that p < 0.05

Lastly, Table 21's analysis of the Overall ASCOPL score highlighted that the High group, with an average score of 4.39, significantly surpassed the Low group, which had an average score of 3.71 (difference of 0.68, p < 0.05), and the Medium group, with an average of 3.77 (difference of 0.61). This indicates that students who maintain a more positive overall attitude towards programming education tend to achieve higher scores on PAT, emphasizing the significance of cultivating comprehensive positive attitudes towards programming for enhancing learning outcomes.

Table 21. Pairwise comparisons of overall scores among different groups for ASCOPL and PAT

| Groups | Mean | Low | Medium | High |
|---|---|---|---|---|
| Low | 3.71 | -- | -0.07 | -0.68* |
| Medium | 3.77 | 0.07 | -- | -0.61* |
| High | 4.39 | 0.68* | 0.61* | -- |

* denoted that p < 0.05

These detailed analyses from Tables 18 to 21 elucidate the significant relationship between various dimensions of ASCOPL and students' programming achievements. They underscore that not only individual attitudes like willingness, negativity, and the perceived necessity of programming are pivotal, but a positive overall attitude towards programming also plays a crucial role in academic success. These insights advocate for educational strategies that enhance specific aspects of students' attitudes towards programming and their overall learning experience to improve academic outcomes in programming education.

**Discussion**

In the discussion of our findings, it was evident that implementing Code.org within a primary school had yielded significant impacts across several dimensions: computational thinking skills, learning motivation, attitudes toward learning computer programming, and student achievement in block-based programming. The main findings of our study are shown in Table 22.

Table 22. Impact of code.org on computational thinking, motivation, attitude, and achievement

| Section | Main Findings |
|---|---|
| PCTS Paired Sample T-test | Significant improvements in Computational Concepts, Practices, and Perspectives from pre-test to post-test, demonstrating the effectiveness of Code.org in enhancing computational thinking skills. |
| IMMS Paired Sample T-test | Notable increases in Attention, Relevance, Confidence, and Satisfaction from pre-test to post-test, indicating the instructional material's positive impact on motivational aspects of learning. |
| ASCOPL Paired Sample T-test | Significant enhancement in Willingness, Negativity, and Necessity dimensions from pre-test to post-test, suggesting improvements in attitudes toward learning computer programming using Code.org. |

| Student Achievement | High completion rates across most Code.org lessons, indicating effective engagement and learning. The analysis also highlighted areas needing additional support, particularly with more advanced concepts. |
|---|---|
| Pearson Correlation Analysis | Strong correlations between ASCOPL, IMMS, PCTS, and student achievement in Code.org, emphasizing the interconnectedness of motivational aspects, attitudes towards programming, and computational thinking skills in influencing learning outcomes. |
| ANOVA Analysis for PCTS and PAT | Statistically significant differences were found in Computational Concepts and Perspectives among groups, underscoring the importance of these domains in programming achievement. No significant variation in Computational Practices, suggesting uniformity in this domain across groups. |
| ANOVA Analysis for IMMS and PAT | Significant differences in Attention, Relevance, and Confidence dimensions among groups, highlighting the role of motivational aspects in programming achievement. There was no significant difference in Satisfaction, indicating similar satisfaction levels across groups. |

Our study made a notable contribution to the existing literature by delving deep into the interconnectedness of computational thinking, learning motivation, learning attitude, and their collective impact on programming achievement within block-based programming. Prior literature had seldom explored these relationships in depth, particularly concerning how they manifested and influenced outcomes in primary education settings. Our research filled a significant gap by examining these elements in concert. It provided insights into the symbiotic relationship between these domains and their role in fostering a conducive learning environment for programming education.

The paired sample T-test results for the PCTS, IMMS, and ASCOPL scales affirmed the hypothesis that targeted instructional materials and approaches could significantly enhance computational thinking skills, motivate students, and cultivate positive attitudes toward programming. Notably, the substantial improvements across the Computational Concepts, Practices, and Perspectives domains highlighted the curriculum's capacity to enhance foundational computational thinking skills, which are critical for programming proficiency.

Moreover, the strong correlations identified between PCTS, IMMS, ASCOPL, and student achievement in Code.org emphasized the integral role of motivational factors, attitudes towards programming, and computational thinking skills in influencing learning outcomes. These findings suggested that fostering a positive learning environment, characterized by engaging instructional materials, a curriculum that resonated with students' interests and goals, and activities that built confidence and satisfaction, enhanced students' programming achievements.

The ANOVA analyses further delineated the relationship between these domains and students' programming achievement, distinguishing the differences among high, medium, and low performers. Significant differences observed in Computational Concepts and Perspectives among groups underscored the pivotal role of these domains in programming achievement. This suggested that a deeper understanding of computational concepts and a positive perspective toward programming were instrumental in achieving higher proficiency levels. Interestingly, the lack of significant variation in Computational Practices suggested a uniformity in this domain across groups, indicating that the Code.org curriculum effectively imparted computational practices to all students regardless of their achievement level.

Similarly, the significant differences in the Attention, Relevance, and Confidence dimensions among groups, as revealed through the ANOVA analysis for IMMS and PAT, highlighted the influence of motivational aspects on programming achievement. This finding pointed to the importance of designing instructional materials and pedagogical strategies that captured students' attention, demonstrated the relevance of programming to their lives, and built their confidence in their programming abilities.

Our research not only corroborated the importance of integrating computational thinking, motivation, and attitudes toward programming in educational strategies but also illuminated the differential impacts of these domains on programming achievement across varied performance groups. By identifying and analyzing these nuanced relationships, our study offered valuable insights for educators, curriculum developers, and policymakers aiming to enhance programming education, mainly through block-based programming platforms like Code.org.

**Conclusion**

This investigation into the effectiveness of Code.org in a primary school setting has yielded significant insights into enhancing computational thinking, motivation, and attitudes toward learning computer programming. The study's findings illuminate the pivotal role of an integrated educational approach that combines rigorous instructional strategies with motivating and engaging materials alongside fostering a positive learning environment for students.

Through a comprehensive analysis, including paired sample t-tests, ANOVA analysis, and correlation assessments, our research demonstrated substantial improvements in students' computational thinking skills, which are crucial for navigating the complexities of the digital age. These skills, encompassing computational concepts, practices, and perspectives, were significantly enhanced, reflecting the effectiveness of the Code.org platform and the instructional approach employed.

Moreover, the study highlighted the importance of learning motivation and attitudes toward programming. The instructional materials used in this study were instrumental in elevating students' engagement and enthusiasm for learning programming, as evidenced by notable increases in attention, relevance, confidence, and satisfaction. These motivational aspects and a positive attitude shift towards computer programming significantly impacted students' overall achievement in programming tasks on Code.org.

The inclusion of ANOVA analysis further enriched our understanding by revealing significant differences among various performance groups, emphasizing the differential impact of computational thinking, motivation, and attitudes on programming achievement. This analysis underscored the need for tailored educational strategies catering to diverse learner needs and capabilities, highlighting the complexity of teaching and learning programming in primary education.

In conclusion, the findings from this study advocate for an educational approach that is not only focused on skill development but also emphasizes the importance of motivation and attitudes toward learning. Such an approach can significantly enhance programming education outcomes, preparing students more effectively for the challenges and opportunities of the digital world. The insights gained from this study offer valuable guidance for educators, curriculum developers, and policymakers in designing and implementing programming education that is engaging, motivating, and effective.

**Reference**

Agapito, J. L., Rodrigo, M., & Mercedes, T. (2017). *Designing an intervention for novice programmers based on meaningful gamification: An expert evaluation*.

Aldrich, C. (2009). *Learning online with games, simulations, and virtual worlds: Strategies for online instruction*. John Wiley & Sons.

Bandura, A., Freeman, W. H., & Lightsey, R. (1999). *Self-efficacy: The exercise of control*. Springer.

Barr, V., & Stephenson, C. (2011). Bringing computational thinking to K-12: What is involved and what is the role of the computer science education community? *Acm Inroads*, *2*(1), 48–54.

Bayliss, J. D., & Strout, S. (2006). Games as a" flavor" of CS1. *Proceedings of the 37th SIGCSE Technical Symposium on Computer Science Education*, 500–504.

Bers, M. U., Flannery, L., Kazakoff, E. R., & Sullivan, A. (2014). Computational thinking and tinkering: Exploration of an early childhood robotics curriculum. *Computers & Education*, *72*, 145–157.

Brennan, K., & Resnick, M. (2012). New frameworks for studying and assessing the development of computational thinking. *Proceedings of the 2012 Annual*

*Meeting of the American Educational Research Association, Vancouver, Canada*, *1*, 25.

Büyüköztürk, Ş., Kılıç Çakmak, E., Akgün, Ö. E., Karadeniz, Ş., & Demirel, F. (2008). *Scientific research methods* (3rd ed.). Ankara: PegemA.

Chang, C. H. (1996). *Educational Psychology-Theories and Practices of Triadic Orientation*. Tung Wah Publishing House.

Chen, G., Shen, J., Barth-Cohen, L., Jiang, S., Huang, X., & Eltoukhy, M. (2017). Assessing elementary students' computational thinking in everyday reasoning and robotics programming. *Computers & Education*, *109*, 162–175.

Erodogan, Y., Aydin, E., & Kabaca, T. (2008). Exploring the psychological predictors of programming achievement. *Journal of Instructional Psychology*, *35*(3), 264.

Felleisen, M., Findler, R. B., Flatt, M., & Krishnamurthi, S. (2004). The TeachScheme! Project: Computing and programming for every student. *Computer Science Education*, *14*(1), 55–77.

Gee, J. P. (2003). What video games have to teach us about learning and literacy. *Computers in Entertainment (CIE)*, *1*(1), 20–20.

Grover, S., Pea, R., & Cooper, S. (2015). Designing for deeper learning in a blended computer science course for middle school students. *Computer Science Education*, *25*(2), 199–237.

Huang, B., & Hew, K. F. T. (2016). Measuring learners' motivation level in massive open online courses. *International Journal of Information and Education Technology*.

Hwang, G.-J., Wu, P.-H., & Chen, C.-C. (2012). An online game approach for improving students' learning performance in web-based problem-solving activities. *Computers & Education*, *59*(4), 1246–1256.

Kalelioğlu, F., Gülbahar, Y., Akçay, S., & Doğan, D. (2014). Curriculum integration ideas for improving the computational thinking skills of learners through programming via scratch. *Local Proceedings of the 7th International Conference on Informatics in Schools: Situation, Evolution and Perspectives*, 101–112.

Kazimoglu, C. (2020). Enhancing confidence in using computational thinking skills via playing a serious game: A case study to increase motivation in learning computer programming. *IEEE Access*, *8*, 221831–221851.

Keller, J. M. (1979). Motivation and instructional design: A theoretical perspective. *Journal of Instructional Development*, 26–34.

Keller, J. M. (2010). The Arcs model of motivational design. *Motivational Design for Learning and Performance: The ARCS Model Approach*, 43–74.

Kim, J. A., & Kim, H. J. (2017). Flipped learning of scratch programming with code. Org. *Proceedings of the 2017 9th International Conference on Education Technology and Computers*, 68–72.

Korkmaz, Ö., & Altun, H. (2014). A Validity and Reliability Study of the Attitude Scale of Computer Programming Learning (ASCOPL). *Online Submission*, *4*(1), 30–43.

Lai, C., Wang, Q., & Lei, J. (2012). What factors predict undergraduate students' use of technology for learning? A case from Hong Kong. *Computers & Education*, *59*(2), 569–579.

Law, K. M., Geng, S., & Li, T. (2019). Student enrollment, motivation and learning performance in a blended learning environment: The mediating effects of social, teaching, and cognitive presence. *Computers & Education*, *136*, 1–12.

Li, M. T. (2014). *A study of first-year polytechnic students' learning attitudes, homework attitudes and academic performance in mathematics*. National Taiwan Normal University.

Liaw, S. S. (2008). Investigating students' perceived satisfaction, behavioral intention, and effectiveness of e-learning: A case study of the Blackboard system. *Computers & Education*, *51*(2), 864–873.

Malan, D. J., & Leitner, H. H. (2007). Scratch for budding computer scientists. *ACM Sigcse Bulletin*, *39*(1), 223–227.

Maryono, D., Budiyono, S., & Akhyar, M. (2022). Implementation of Gamification in Programming Learning: Literature Review. *Int. J. Inf. Educ. Technol*.

Mirvis, P. H. (1991). *Flow: The psychology of optimal experience*. JSTOR.

Myers, B. A. (1986). Visual programming, programming by example, and program visualization: A taxonomy. *ACM Sigchi Bulletin*, *17*(4), 59–66.

Myers, D. G. (1993). Behavior and attitudes. *Social Psychology:*

Ngan, L. C. (1980). A study of university students' attitude towards Chinese language. *Journal of the College of Education*, *5*, 3–125.

Noone, M., & Mooney, A. (2018). Visual and textual programming languages: A systematic review of the literature. *Journal of Computers in Education*, *5*, 149–174.

Plass, J. L., Homer, B. D., & Kinzer, C. K. (2015). Foundations of game-based learning. *Educational Psychologist*, *50*(4), 258–283.

Pugnali, A., Sullivan, A., & Bers, M. U. (2017). The impact of user interface on young children's computational thinking. *Journal of Information Technology Education. Innovations in Practice*, *16*, 171.

Román-González, M., Pérez-González, J.-C., & Jiménez-Fernández, C. (2017). Which cognitive abilities underlie computational thinking? Criterion validity of the Computational Thinking Test. *Computers in Human Behavior*, *72*, 678–691.

Selby, C., & Woollard, J. (2013). *Computational thinking: The developing definition*.

Tan, P.-H., Ting, C.-Y., & Ling, S.-W. (2009). Learning difficulties in programming courses: Undergraduates' perspective and perception. *2009 International Conference on Computer Technology and Development*, *1*, 42–46.

Tang, X., Yin, Y., Lin, Q., Hadad, R., & Zhai, X. (2020). Assessing computational thinking: A systematic review of empirical studies. *Computers & Education*, *148*, 103798.

Tsou, C. Y. (2014). *The effect on computer technology learning by means of PBL strategy to fifth grade students：Taking the learnong of Scratch program language as an example*. NTOU-National Taiwan Ocean University.

Wang, K. X. (1996). *Psychology of learning*. Laureate Book Company.

Wang, W. K., Shao, R. Z., & Pi, L. S. (1995). *Educational Psychology*. Wunan Publishing Company.

Wang, X.-M., Hwang, G.-J., Liang, Z.-Y., & Wang, H.-Y. (2017). Enhancing students' computer programming performances, critical thinking awareness and attitudes towards programming: An online peer-assessment attempt. *Journal of Educational Technology & Society*, *20*(4), 58–68.

Weintrop, D. (2019). Block-based programming in computer science education. *Communications of the ACM*, *62*(8), 22–25.

Weintrop, D., Beheshti, E., Horn, M., Orton, K., Jona, K., Trouille, L., & Wilensky, U. (2016). Defining computational thinking for mathematics and science classrooms. *Journal of Science Education and Technology*, *25*, 127–147.

Werner, L., Denner, J., Campe, S., & Kawamoto, D. C. (2012). The fairy performance assessment: Measuring computational thinking in middle school. *Proceedings of the 43rd ACM Technical Symposium on Computer Science Education*, 215–220.

Wing, J. M. (2006). Computational thinking. *Communications of the ACM*, *49*(3), 33–35.

Wing, J. M. (2008). Computational thinking and thinking about computing. *Philosophical Transactions of the Royal Society A: Mathematical, Physical and Engineering Sciences*, *366*(1881), 3717–3725.

Wu, H. W. (2013). *The Influences of International Education on the English Learning Attitudesand the English Learning Effects of Senior High School Students-a Case of National Huwei Senior High School*. Asia University.

Wu, P. Z. (2021). *Compilation of Computational Thinking Self-efficacy Scale and Verification of Reliability and Validity*. Taichung University of Education.

Yang, J., Wong, G. K., & Dawes, C. (2018). An exploratory study on learning attitude in computer programming for the twenty-first century. *New Media for Educational Change: Selected Papers from HKAECT 2018 International Conference*, 59–70.

Zhong, B., Wang, Q., Chen, J., & Li, Y. (2016). An Exploration of Three-Dimensional Integrated Assessment for Computational Thinking. *Journal of Educational Computing Research*, *53*(4), 562–590. https://doi.org/10.1177/0735633115608444

Zhu, J. X. (1986). *Psychology of learning*. Chihuahua Publishing Company.